\documentclass[aps,prl,twocolumn,raggedbottom,superscriptaddress,reprint]{revtex4-2}

\usepackage[dvipsnames]{xcolor}
\usepackage{amssymb,amsmath,amsthm,amsfonts} 
\usepackage{mathtools,mathrsfs}
\usepackage[cal=pxtx]{mathalpha}
\usepackage{hyperref}
\usepackage{cleveref}
\usepackage{booktabs}
\usepackage{braket}
\usepackage{comment}

\usepackage{graphicx,subcaption,tikz}

\definecolor{apslinkblue}{RGB}{50,48,142}
\hypersetup{
    pdfencoding=unicode,
	colorlinks=true,
	urlcolor=apslinkblue,
	linkcolor=apslinkblue,
	citecolor=apslinkblue,
	pdftitle={Worldsheet Duals to One-Matrix Models},
	pdfauthor={Giacchetto Alessandro, Gopakumar Rajesh, Mazenc Edward A.},
	pdfdisplaydoctitle=true,
	pdfstartview=FitH,
	linktocpage=true
}

\newcommand{\cO}{\mathcal{O}}
\newcommand{\cV}{\mathcal{V}}
\newcommand{\cS}{\mathcal{S}}
\newcommand{\cT}{\mathcal{T}}
\newcommand{\cN}{\mathcal{N}}
\newcommand{\SU}{\mathrm{SU}}
\newcommand{\U}{\mathrm{U}}

\newcommand{\M}{\mathcal{M}}
\newcommand{\Mbar}{\overline{\mathcal{M}}}

\DeclareMathOperator{\Tr}{Tr}
\DeclareMathOperator*{\Res}{Res}
\newcommand{\Crit}{\mathrm{Crit}}
\newcommand{\bigO}{\mathrm{O}}
\newcommand{\re}{\mathrm{e}}

\DeclarePairedDelimiter\ceil{\lceil}{\rceil}
\DeclarePairedDelimiter\floor{\lfloor}{\rfloor}

\newcommand{\corr}[3]{\langle \thinspace {#1} \thinspace \rangle_{#2}^{\textup{#3}}}
\newcommand{\Corr}[3]{\left\langle {#1} \right\rangle_{\!\! #2}^{\!\!\textup{#3}}}
\newcommand{\corrint}[3]{\langle\!\!\!\langle \thinspace {#1} \thinspace \rangle\!\!\!\rangle_{#2}^{\textup{#3}}}
\newcommand{\Corrint}[3]{\left\langle\!\!\!\!\left\langle {#1} \right\rangle\!\!\!\!\right\rangle_{\!\! #2}^{\!\!\textup{#3}}}

\begin{document}
\title{Worldsheet Duals to One-Matrix Models}
\author{Alessandro Giacchetto}
\affiliation{Departement Mathematik, ETH Zürich, 8006 Zürich, Switzerland}
\author{Rajesh Gopakumar}
\affiliation{International Centre for Theoretical Sciences-TIFR, Shivakote, Hesaraghatta Hobli, Bengaluru North 560089, India}
\author{Edward A. Mazenc}
\affiliation{Institut für Theoretische Physik, ETH Zürich, 8093 Zürich, Switzerland}
\begin{abstract}
   We derive a concrete closed string dual to any interacting Hermitian one-matrix model, away from the double-scaling limit. Matrix and string correlators manifestly agree, to all orders in the genus expansion and all orders in the 't~Hooft coupling(s). The worldsheet theory consists of a supersymmetric B-twisted Landau--Ginzburg model coupled to 2d topological gravity. We provide a precise dictionary between traces of the matrix and vertex operators on the worldsheet. Matrix model correlators are explicitly mapped to computable integrals over the moduli space of Riemann surfaces. We perform several direct cross-checks on both sides of the duality. This work furnishes a detailed instantiation of gauge/string duality, in the standard 't~Hooft regime, and hopefully a useful worldsheet toy model for the AdS/CFT correspondence, away from the free field limit.
\end{abstract}

\maketitle

\section{\label{sec:into}Introduction}  
Gauge/string duality relates the dynamics of large-$N$ matrix-valued fields to that of 2d surfaces, the dual string worldsheets. It underpins the holographic correspondence \cite{Mal98,GKP98,Wit98}, which has transformed theoretical physics over the past three decades. Historically, a string description of $\SU(N)$ Yang--Mills was first sought as a reformulation that would naturally access the strong-coupling regime $g_{\rm YM}^2 = \bigO(1)$, where ordinary perturbation theory breaks down. To this end, 't~Hooft introduced the alternative small parameter $1/N$, with $N$ the rank of the gauge group \cite{tHo74}. In the large-$N$ limit, keeping $\lambda = g_{\rm YM}^2 N$ fixed, gauge-theory Feynman diagrams are organized by topology, mirroring the genus expansion of string theory with $g_{\rm s} \sim 1/N$. This was suggestive of a dual string description that would compute gauge-theory observables order by order in $1/N$, by considering worldsheets of fixed genus, while re-summing the expansion in the coupling $\lambda$ into a parameter of the target geometry. In this letter, we carry out this program explicitly when the gauge theory reduces to a finite-dimensional integral over Hermitian $N \times N$ matrices. Crucially, the dual worldsheet theory remains tractable for finite 't~Hooft couplings. 

Matrix integrals were already used in \cite{BIPZ78} as toy models to illustrate the 't~Hooft limit in practice, resumming planar (genus zero) graphs via saddle-point techniques. Yet, in the half-century since, the second part of 't~Hooft's program---namely, deriving a concrete worldsheet string description for these models---has remained incomplete.
A major obstacle has been that continuum string theories were long thought to emerge only in the double-scaling limit of matrix integrals, rather than in the standard 't~Hooft limit. In that regime, the 't~Hooft couplings are tuned to critical values and the genus expansion parameter is no longer simply $1/N$. The resulting dual string theories, such as 2d minimal models coupled to Liouville gravity, therefore have no tunable parameters to encode the full coupling dependence of matrix-model observables. Double-scaled models thus differ sharply from more familiar instances of gauge/string duality, such as the AdS/CFT correspondence. 

Our construction is also, \textit{prima facie}, distinct from earlier proposals of B-model topological strings on certain non-compact Calabi--Yau 3-folds as duals of matrix integrals away from the double-scaling limit \cite{DV02a,DV02b,Mar06,ADKMV06,DV07,PvHV22}. Moreover, in the latter context, an explicit operator dictionary and direct checks at the level of worldsheet observables have remained difficult \footnote{
    Proposals for an A-model string dual have also been made. See \cite{Gop11, GP13, GKKMS26, GM, GGLM} for this line of development which ought to be mirror to the present proposal.
}. Yet, this is precisely the core of gauge/string duality: the equality between gauge theory correlators of single-trace operators and independently defined worldsheet amplitudes, computed as integrals over moduli spaces of punctured Riemann surfaces. As emphasized recently in \cite{Gai26}, there has been no systematic way to derive this integrand on moduli space, even for, say, the quartic matrix model. This letter aims to remedy this situation.

\section{\label{sec:main}Main Result}
We rewrite the connected genus-$g$, $n$-point single-trace correlator in any one-matrix model as a computable integral over the moduli space of Riemann surfaces $\Mbar_{g,n}$:
\begin{equation} \label{eq:main:result}
    \Corr{
        \prod_{i=1}^n \Tr{M^{k_i}}
    }{g}{MM}
    =
    \int_{\Mbar_{g,n}}
        \Corrint{ \prod_{i=1}^n \cV_{k_i} }{g}{LG+grav} ,
\end{equation}
where the left-hand side denotes the $\bigO(N^{2-2g-n})$ order term in the large-$N$ expansion of the matrix-model correlator, while on the right-hand side $g$ denotes the genus of the worldsheet.
The string integrand is computable for finite 't~Hooft coupling(s), thus realizing the original hope for a dual worldsheet description. Moreover, the integral is taken over the Deligne--Mumford compactification $\Mbar_{g,n}$; boundary contributions from degenerate worldsheets play a crucial role. Most importantly, the integrand arises from an explicit worldsheet theory: a B-twisted $\cN = 2$ Landau--Ginzburg (LG) model with a single chiral superfield, coupled to 2d topological gravity \cite{Vaf90,Wit91,DVV90}.

As summarized in \cref{eq:correspond}, each large-$N$ saddle of the matrix integral, specified by its spectral curve, determines the target geometry of the dual string. This curve carries two functions, $x$ and $y$; in the worldsheet theory, $x$ becomes the LG superpotential and $dy$ the Calabi--Yau top form. The resulting LG model computes the corresponding $1/N$ expansion of matrix-model observables around that saddle.

In fact, the equivalence is even stronger: \cref{eq:main:result} follows from an explicit operator dictionary between single-trace and worldsheet vertex operators. On the worldsheet, these operators are built from matter primaries $\mathcal O_\alpha$ (elements of the LG chiral ring) and gravitational descendants $\psi^k$ from the topological gravity sector. In a large-$N$ saddle where the eigenvalue distribution is supported on $\mathfrak c$ intervals, the dual string has $2\mathfrak c$ matter primaries. For concreteness, in the one-cut phase ($\mathfrak c=1$) the resulting quasi-universal dictionary, valid for all interacting one-matrix models, reads
\begin{equation}\label{eq:op:dict}
    \Tr{ M^k }
    \leftrightarrow
    \cV_k
    \coloneqq\!
    \biggl[
        \frac{k!\gamma u e^{\delta u}}{1 - u\psi}
        \sum_{\alpha = \pm 1}
            \cO_{\alpha}
            \frac{I_{0}(2 \gamma u)+ \alpha I_{1}(2 \gamma u)}{\Omega_\alpha}
    \biggr]_k,
\end{equation}
where $[f(u)]_{k}$ extracts the $k$-th coefficient of the Taylor expansion of $f(u)$ around $u = 0$, while $I_0$ and $I_1$ are modified Bessel functions. The dependence on the matrix model potential enters only through four geometric constants: $\gamma$, $\delta$, and $\Omega_\pm$. The parameters $\gamma$ and $\delta$ encode the width and center of the eigenvalue support, while $\Omega_\pm$ control its square-root behavior at the endpoints.

What is remarkable about this string theory is that we can explicitly evaluate the correlators of the vertex operators $\cV_k$ for finite values of the 't~Hooft couplings. This should be contrasted with the, obviously much more complicated, AdS/CFT correspondence, 
where bona fide worldsheet calculations largely remain intractable, even at genus zero.

The much simpler topological string theories considered here are solvable because they admit a reformulation in the mathematical language of (semisimple) Cohomological Field Theory (CohFT) \cite{KM94,Giv01,Tel12}. This framework also underlies the recently discovered Virasoro and Complex Liouville string dualities \cite{CEMR23,CEMR24}. In a nutshell, it provides a concrete algorithm for constructing the integrand on moduli space for the matter theory coupled to 2d topological gravity. Mathematically, this consists of three basic ingredients, namely, the \textit{translation} $T$, the \textit{rotation} $R$, and the data of a topological field theory. Physically, they correspond to certain worldsheet disk partition functions, together with the pure matter correlators of the LG model. Identifying these quantities directly from the worldsheet theory---see \cref{eq:T,eq:R}---is one of the central results of this and upcoming work \cite{GGM}. In the end, CohFT allows us to write explicit expressions for $\corrint{\prod_{i=1}^{n}\cV_{k_{i}}}{g}{LG+grav}$ in terms of well-studied cohomology classes on $\Mbar_{g,n}$, whose integrals are known. For a given matrix potential, the attached computer implementation constructs the relevant class on moduli space and evaluates its integral \cite{code}. We illustrate this in two examples in the \hyperref[sec:checks]{Explicit Cross-Checks} section, discussed in more detail in the \hyperref[sec:supp]{Supplementary Material}. The computation of matrix-model observables is thereby reduced to intersection theory on moduli space. Such a reduction to algebraic geometry was famously achieved by Witten for the duals of double-scaled matrix models \cite{Wit91,Wit92}. In this sense, our work extends those ideas to arbitrary one-matrix models in the standard 't~Hooft limit. 

Why do the matrix and string correlators agree? In short, both sides obey the same recursion relations. On the matrix-model side, this is the topological recursion on the spectral curve \cite{Eyn04,CEO06,EO07}. On the worldsheet side, they arise from the universal behavior of the theory under worldsheet degenerations \cite{Wit91,VV91}. Our derivation of the operator dictionary is arranged so that the initial data of the two recursive constructions agree to all orders in the 't~Hooft couplings. The equality of all higher-genus correlators then follows recursively, order by order in $1/N$.

\section{\label{sec:matrix}The Matrix Integrals}
Consider the finite-dimensional integrals over Hermitian $N\times N$ matrices, with polynomial potential $V(M) = \frac{1}{2} M^2 - \sum_{k\geq 3}^{d} \frac{t_k}{k}M^k$:
\begin{equation}
    Z_N(t_3,\ldots,t_d)
    =
    \int dM \ \re^{- N \Tr V(M)}
    .
\end{equation}
We study this model in the large-$N$ limit, keeping the 't~Hooft couplings $t_k$ fixed. It may be viewed as a $(0+0)$-dimensional gauge theory, invariant under $M\mapsto U^\dagger M U$ for $U\in \U(N)$, with gauge-invariant observables generated by traces of powers of $M$.
A standard Schwinger--Dyson argument shows that the planar resolvent satisfies an algebraic equation:
\begin{equation}
    y^2 -V'(x)y + P(x)=0,
    \quad
    y(x) = \Corr{\Tr \frac{1}{x-M} }{g=0}{}
,
\end{equation}
where $P(x)$ is a polynomial of degree $d-2$ determined by the choice of large-$N$ saddle. This equation defines a complex curve $\cS$, the spectral curve of the matrix model.

In the one-cut phase, $\cS \cong \mathbb{CP}^1$, and one may introduce a global uniformizing coordinate $z$ in which
\begin{equation} \label{eq:spec:curve}
    x = \gamma \left( z+\frac{1}{z} \right) + \delta,
    \qquad
    y= \sum_{k=1}^{d-1} u_k z^k.
\end{equation}
The constants $\gamma$ and $\delta$ encode the endpoints of the eigenvalue support, while the coefficients $u_k$ are fixed by the potential through $V'(x(z)) = \sum_k u_k(z^k + z^{-k})$. While $y$ depends on the detailed form of the potential, the function $x$ is quasi-universal: it always has two ramification points, at $z = \pm 1$, where $dx=0$.

Even away from the one-cut phase, the spectral curve is the starting point for the topological recursion formalism \cite{EO07}, which recursively computes the symmetric differentials $\omega_{g,n}$ encoding the genus-$g$, $n$-point correlators via a residue calculus at the ramification points of $x$:
\begin{equation} \label{eq:resolvents}
    \omega_{g,n}(z_1,\ldots,z_n)
    = \!
    \Corr{\prod_{i=1}^n \Tr \frac{dx_i}{x_i-M}}{g}{} \!
    +
    \delta_{g,0}\delta_{n,2}
    \frac{dx_1 dx_2}{(x_1 -x_2)^2}.
\end{equation}
In particular, $\omega_{0,2}$ is the Bergman kernel of the spectral curve and will play an important role later.

\section{\label{sec:string}The Dual String Theories}
\textbf{The Matter Sector.}
Topological LG models were introduced in \cite{Vaf90}, in search of a Lagrangian description of the chiral ring of certain superconformal field theories \cite{VW89,LVW89}. They often arise by B-twisting an $\cN = 2$ sigma model with non-compact Calabi--Yau target $\cT$ and holomorphic superpotential $W$, which we assume has isolated non-degenerate critical points \cite{LL91}.
For a one-complex-dimensional target \footnote{
    We will assume that $\cT$ is an affine algebraic curve, i.e. the complement of finitely many points in a compact Riemann surface; this will guarantee the existence of a canonical Bergman kernel for fixed $\mathcal{A}$ and $\mathcal{B}$ cycles.
}, the field content is simple: a bosonic map $\phi \colon \Sigma \to \cT$, fermionic scalars $\chi, \bar\chi \in \Gamma(\Sigma,\phi^*\bar T\cT)$, and a fermionic one-form $\rho \in \Gamma(\Sigma,\Omega^1_\Sigma \otimes \phi^*T\cT)$.
After integrating out auxiliary fields, the twisted action on a genus-$g$ worldsheet $\Sigma$ with metric $h_{\mu\nu}$ is
\begin{multline}\label{eq:LG:action}
  S
  =
  \int_\Sigma
    d^2\sigma \sqrt{h} \Big(
        h^{\mu\nu} G_{\phi \bar\phi}
        \partial_\mu \phi \,
        \partial_\nu \bar\phi
        + \frac{1}{8} \,
        G^{\phi \bar{\phi}}
        (\partial_{\bar\phi}\bar W)
        (\partial_\phi W)
        \\
        - \mathrm{i} \,
        G_{\phi \bar\phi}
        \chi D_{\bar\sigma} \rho_{\sigma}
        + \mathrm{i} \,
        G_{\phi\bar\phi}
        \bar\chi D_{\sigma} \rho_{\bar\sigma}
        \\
        +\frac{1}{4} \,
        (\nabla_\phi \partial_\phi W) 
        \rho_{\bar\sigma} \rho_{\sigma}
        + \frac{1}{4} \,
        (\nabla_{\bar\phi}\partial_{\bar\phi}\bar W)\chi \bar\chi
    \Big).
\end{multline}
Here, $G_{\phi\bar\phi}$ is a Ricci-flat Kähler metric on the target, although physical observables will not depend on this choice \cite{Wit91b,LL91,LM94}. The covariant derivatives $D$ and $\nabla$ are the pullbacks of target-space derivatives to the worldsheet; their explicit form will not matter below (see, e.g., \cite{Hor03,Oog12}). The model possesses a nilpotent scalar fermionic symmetry $Q_{\rm B}$. Although the action is not $Q_{\rm B}$-exact, the stress tensor is. So correlators of $Q_{\rm B}$-closed observables are topological and, in particular, independent of the insertion points.

Crucially, the LG path integral can be evaluated exactly \cite{Vaf90}: it is semiclassically exact and localizes onto constant maps $\phi(\sigma) = \alpha$ where $dW(\alpha) = 0$. In other words, the worldsheet collapses onto the critical points $\alpha \in \Crit(W)$ of the superpotential. For non-zero modes, bosonic and fermionic determinants cancel by supersymmetry. Zero modes require a choice of holomorphic nowhere-vanishing top form $\Omega$ on the target, a.k.a. a Calabi--Yau form; this \emph{additional choice} is part of the definition of the theory. The resulting zero-mode integrals produce genus-dependent powers of the Hessian $H_\alpha \coloneqq \frac{\nabla dW}{\Omega^2} \big|_\alpha$ at each critical point \footnote{
    At a critical point, $\nabla dW$ is intrinsically a quadratic holomorphic form, independent of the choice of connection; since $\Omega$ trivializes the canonical bundle $K_{\cT}$, the ratio $(\nabla dW)/\Omega^2$ is a well-defined scalar.
}.

This makes it possible to evaluate not only the partition function, but more generally all correlation functions. The $Q_{\rm B}$-closed operators form a ring, isomorphic to the chiral ring of the untwisted theory. It is generated by holomorphic functions on $\cT$, modulo the relation $dW=0$. Their exact genus-$g$ correlators are
\begin{equation}
\begin{split}
    \Corr{
        \prod_{i=1}^{n} F_{i}
    }{g}{LG}
    &=
    \sum_{\alpha \in \Crit(W)}
        \left( \prod_{i=1}^n F_{i}(\alpha) \right)
        H_\alpha^{g-1}
    \\
    &=
    \sum_{\alpha \in \Crit(W)}
    \Res_{\alpha}
        \left( \prod_{i=1}^n F_{i} \right)
            H^{g} \frac{\Omega^2}{dW} .
\end{split}
\end{equation}
Thus, the answer can be written as a residue at the critical points of $W$, with $H^g$ playing the role of the handle operator in this topological field theory.

A final fact will be crucial below. Since the critical points are isolated and non-degenerate, the chiral ring is \emph{semisimple}, meaning that it admits a basis of $Q_{\rm B}$-closed operators $\cO_\alpha$ that are idempotent: $\cO_\alpha \cdot \cO_\beta = \delta_{\alpha,\beta} \, \cO_\alpha$. Each such operator is supported on a single critical point of $W$. In this basis, the correlators are diagonal:
\begin{equation} \label{eq:LG:diag:corr}
    \Corr{
        \prod_{i=1}^{n} \cO_{\alpha_i}
    }{g}{LG}
    =
    \sum_{\alpha \in \Crit(W)} \delta_{\alpha_1,\ldots,\alpha_n,\alpha} \, H_\alpha^{g-1}.
\end{equation}

\textbf{Coupling to Topological Gravity.}
To promote the twisted LG model to a closed string theory, one must couple it to 2d topological gravity on the worldsheet \cite{Wit88,LPW88,MS89,Dis90,Wit90,VV91,DVV91,Dij92,BCI94,GR05}. Various field-theoretic realizations of topological gravity exist \cite{LPW88,MS89,Dis90,VV91, BCI94,GR05}. We follow the algebro-geometric approach of \cite{Wit91,Wit93}. The resulting BRST-closed observables are obtained by dressing matter primaries with gravitational descendants. Concretely, the basic observables of the coupled theory are of the form $\cO_\alpha \psi^k$, where $\psi$ is the cotangent-line class at a marked point of $\Mbar_{g,n}$. Physically, $\psi$-classes represent closed differential two-forms on moduli space.

For each genus $g$ and number $n$ of operator insertions, the gravity-coupled theory produces a canonical integrand on the moduli space $\Mbar_{g,n}$, which we denote by $\corrint{ \prod_i \cO_{\alpha_i}}{g}{LG+grav}$. Its integral over $\Mbar_{g,n}$, denoted with a single bracket $\corr{ \prod_i \cO_{\alpha_i}}{g}{LG+grav}$, gives the corresponding worldsheet amplitude. This integrand is constrained by the usual factorization properties of the worldsheet path integral: whenever a cycle pinches, or operator insertions collide, the amplitude must decompose in a way compatible with sewing. In mathematical language, these integrands define cohomology classes on $\Mbar_{g,n}$ that satisfy the axioms of a Cohomological Field Theory (CohFT) \cite{KM94,GL26} with a choice of vacuum \cite{CGG25}. 

These factorization constraints do not always uniquely determine the theory. However, in the semisimple case relevant here, a fundamental reconstruction theorem \cite{Giv01,Tel12} implies that the full family of integrands $\corrint{\prod_i \cO_{\alpha_i}}{g}{LG+grav}$ is determined by three pieces of data: the pure matter correlators, a vector-valued function $T$, and a matrix-valued function $R$. Here, $T$ governs the string background, while $R$ governs how local operator insertions are represented in the sewing formalism. Concretely, the integrand of the gravity-coupled theory is obtained from the pure LG matter correlator:
\begin{equation}\label{eq:corr:LGgrav}
    \Corrint{
        \prod_{i=1}^n \cO_{\alpha_i}
    }{g}{LG+grav}
    =
    R.T. \Corr{
        \prod_{i=1}^n \cO_{\alpha_i}
    }{g}{LG} .
\end{equation}
In the \hyperref[sec:supp]{Supplementary Material}, we sketch the precise meaning of the right-hand side, along with its physical interpretation. A critical contribution of this work is to identify $T$ and $R$ directly in terms of target-space exponential integrals. We postpone their derivation from the worldsheet path integral to \cite{GGM}, but record the final expressions:
\begin{align}
    \label{eq:T}
    T^\alpha(u)
    &=
    u - \sqrt{ \frac{u H_\alpha}{2\pi} }
    \int_{\Gamma_\alpha} \re^{-\frac{W - W_\alpha}{u}} \, \Omega, \\
    \label{eq:R}
    R^\alpha_\beta(u)
    &=
    \sqrt{ \frac{u H_\alpha}{2\pi} }
    \int_{\Gamma_\alpha} \re^{-\frac{W - W_\alpha}{u}} \, \theta_\beta .
\end{align}
Here, $\Gamma_\alpha$ is the Lefschetz thimble emanating from the critical point $\alpha$, and $W_\alpha$ is the critical value of the superpotential there. The differential $\theta_\beta(\phi) \coloneqq -\frac{1}{H_{\beta }}\frac{B(\phi,\cdot)}{\Omega}|_{\beta}$ is a meromorphic form dual to $\cO_\beta$, defined in terms of a Bergman kernel $B$ on the target. The latter is an \emph{additional choice} in the gravity-coupled theory, equivalent to a choice of $\mathcal{A}$ and $\mathcal{B}$ cycles on the target, which in turn fixes the normalization of all $\theta_\beta$ along such cycles. From the point of view of the worldsheet, $T^{\alpha}(u)$ can be understood as an equivariant disk partition function, with boundary condition labeled $\alpha$. The equivariant parameter $u$ codifies the rotations of the disk. Similarly, $R^\alpha_\beta(u)$ corresponds instead to an equivariant disk partition function with an additional insertion of an operator dual to $\cO_\beta$ at its center. These quantities appeared---for entirely different reasons---in \cite{CV93,Nek18}. Once $T$ and $R$ are known, the full integrand on $\Mbar_{g,n}$ is completely determined.

There exists an alternative mathematical approach to LG models coupled to gravity, built around the theory of higher residues and Frobenius manifolds \cite{Los93,Los94,Los98,Sai83,Dub96,CL12,LLS13}. Although far from the context of gauge/string duality, that formalism has been successfully applied to singularity theory and mirror descriptions of Gromov--Witten theories \cite{FJR13,KS11,MS16,HLSW21,FLZ17,DNOS18,DNOS19,CCGG24,Lys25}.
Rather than starting from those formalisms, which are mostly BCOV-like \cite{BCOV93}, we work directly with the worldsheet theory and recover the relevant data from it. A detailed comparison will appear elsewhere.

\section{\label{sec:corr}The Correspondence}
We are now ready to state the precise identification between matrix-model and worldsheet data, valid for any large-$N$ saddle of any interacting one-matrix model:
\begin{equation} \label{eq:correspond}
    \cT = \cS \setminus \set{ \text{pts} },
    \quad
    W = x,
    \quad
    \Omega = dy,
    \quad
    B = \omega_{0,2}.
\end{equation}
The spectral curve $\cS$ of the matrix model, with the poles of $x$ and the zeros and poles of $dy$ excised, thus becomes the target of the string. The identification $W=x$ explains why both matrix-model and LG computations reduce to residue calculus. The ramification points of $\cS$, where $dx=0$, are precisely the critical points of $W$ onto which the worldsheet localizes. The function $y$, in turn, determines the holomorphic top form $\Omega$, while the Bergman kernel $\omega_{0,2} = B$ fixes the additional target-space cycle normalization. The operator correspondence generalizing \cref{eq:op:dict} beyond the one-cut phase is
\begin{equation}\label{eq:op:dict:gen}
     \Tr{M^k}
     \longleftrightarrow
     \sum_{\substack{\alpha \in \Crit(W) \\ m=0,\ldots,k-1}}
        \cO_\alpha \psi^m
        \Res_{W = \infty}
            \frac{k!}{(k-m)!}
            W^{k-m} \, \theta_\alpha ,
\end{equation}
where we recall the differential forms $\theta_\alpha(\phi) \coloneqq -\frac{1}{H_{\alpha}}\frac{B(\phi,\cdot)}{\Omega}|_{\alpha}$ dual to $\cO_\alpha$. Thus, the map between gauge and string operators reduces to an explicit residue computation.

In the one-cut phase, the superpotential takes the quasi-universal form $W = \gamma(\phi + \phi^{-1}) + \delta$, cf. \cref{eq:spec:curve}. The critical points are at $\phi = \pm 1$, and the ring relation $dW = 0$ reduces to $\phi^2 = 1$. The corresponding idempotents are $\cO_{\pm} = (1 \pm \phi)/2$. Apart from the constants $\gamma$ and $\delta$, all dependence on the matrix potential is encoded in $\Omega=dy$. Moreover, in the one-cut phase the Bergman kernel is uniquely fixed to $B(\phi_1,\phi_2) = \frac{d\phi_1 d\phi_2}{(\phi_1 - \phi_2)^2}$, so the operator dictionary in \labelcref{eq:op:dict:gen} reduces precisely to \cref{eq:op:dict}. In the context of the Gaussian matrix model, this superpotential already appeared through its connection to the $c=1$ string theory, see \cite[fig.~1]{GM}.

Why is this the correct dictionary? Matrix-model correlators can be computed by topological recursion on its spectral curve \cite{Eyn04,CEO06,EO07}. The output of topological recursion can in turn be repackaged as integrals over $\Mbar_{g,n}$ \cite{Eyn14}, and \cite{DOSS14} identifies the corresponding integrand in the language of CohFTs. By deriving \emph{directly from the worldsheet theory} explicit expressions for the pure matter correlators, $T$, and $R$ (\cref{eq:LG:diag:corr,eq:T,eq:R}), and applying the correspondence in \cref{eq:correspond}, we show that the same integrand arises from a bona fide string theory. This makes the gauge/string duality manifest. The operator dictionary follows from interpreting the $\omega_{g,n}$'s either in terms of matrix-model resolvents or CohFT observables. Earlier links between LG theories and topological recursion \cite{FLZ17,DNOS18,DNOS19}, including a target space perspective \cite{IO}, further support this connection.

\section{\label{sec:checks}Explicit Cross-Checks}
We now illustrate how worldsheet correlators reproduce matrix-model ones to all orders in the 't~Hooft coupling(s). We focus on the one-cut saddle of the quartic matrix model, $V(M) = \frac{1}{2}M^2 - \frac{t_4}{4}M^4$, for which $\cS \cong \mathbb{CP}^1$ and
\begin{equation}\label{eq:SC:quart}
    x = \gamma \left( z + z^{-1} \right),
    \qquad
    y = \gamma \bigl(  z - t_4 \gamma^2 ( 3z + z^3 ) \bigr),
\end{equation}
with $\gamma^2 = \frac{1-\sqrt{1-12 t_4}}{6t_4}$ and $\omega_{0,2} = \frac{dz_1 dz_2}{(z_1 - z_2)^2}$.
Intermediate steps are given in the \hyperref[sec:supp]{Supplementary Material}.

\begin{widetext}
\textbf{Genus-0, 3-point.}
We begin with the matrix-model correlators. For the special case of even traces in the strict planar limit, the correlator can be computed to all orders in perturbation theory using the planar $n$-point correlator of even traces for the GUE derived in \cite{GP13}:
\begin{equation}\label{eq:g0:3pt:perturb}
    \Corr{
        \prod_{i=1}^{3} \Tr M^{2k_i}
    }{g=0}{quart}
    =
    \sum_{s=0}^{\infty}
        \frac{1}{s!}
        \Corr{
            \biggl( \frac{N t_4 \Tr M^4}{4} \biggr)^s
            \prod_{i=1}^{3} \Tr M^{2k_i}
        }{g=0}{GUE}
    =
    \sum_{s=0}^{\infty}
        \frac{(3t_{4})^s}{s!}
        \frac{(|k|+2s-1)!}{(|k|+s-1)!}
        \prod_{i=1}^{3} \frac{(2k_i)!}{(k_i-1)!k_i!} .
\end{equation}
Here, $|k| = k_1 + k_2 + k_3$.
On the string side, since $\Mbar_{0,3} = \mathrm{pt}$, there is no moduli-space integral to perform. This case therefore probes the duality purely at the level of the matter sector. Using the fact that the correlators of $\cO_\alpha$ are diagonal, the right-hand side of \cref{eq:main:result} becomes
\begin{equation}\label{eq:g0:3pt:exact}
    \Corr{
        \prod_{i=1}^{3} \cV_{2k_i}
    }{g=0}{quart}
    =
    \left(
        \frac{\corr{\cO_{+}^3}{g=0}{LG}}{\Omega_{+}^3}
        -
        \frac{\corr{\cO_{-}^3}{g=0}{LG}}{\Omega_{-}^3}
    \right)
    \prod_{i=1}^{3} \frac{\gamma^{2k_i}(2k_i)!}{(k_i-1)!k_i!}
    =
    \frac{\gamma^{2|k|-2}}{\sqrt{1-12t_{4}}}
    \prod_{i=1}^{3} \frac{(2k_i)!}{(k_i-1)!k_i!} .
\end{equation}
The last equality follows from a direct evaluation of the LG correlators using the identifications \labelcref{eq:correspond,eq:op:dict:gen}; the intermediate equality in fact holds for any even potential in the one-cut phase. Substituting the expression for $\gamma^2$ and expanding in $t_4$, one recovers \cref{eq:g0:3pt:perturb} exactly. Hence, the worldsheet theory resums the full perturbative expansion of the planar three-point function.
\end{widetext}

\textbf{Genus-1, 1-point.}
In genus one, no closed formula is available for the $n$-point GUE correlators. As such, no analog of the perturbative computation in \labelcref{eq:g0:3pt:perturb} can be carried out.
It is also the first example in which the topological gravity sector contributes non-trivially. In particular, it provides a direct test of the CohFT formalism, which expresses $\corrint{\cV}{g=1}{LG+grav}$ as a sum of explicit classes on moduli space.
As explained in the \hyperref[sec:supp]{Supplementary Material}, the final answer reduces to three contributions:
\begin{multline}
    \corr{\cV_{2k}}{g=1}{quart}
    =
    \int_{\Mbar_{1,1}} \Big(
        \frac{16 k-1}{8 \gamma} \psi_1 
        -\frac{3(1-38 t_{4}\gamma^2)}{8\gamma(1-6t_{4}\gamma^2)} \kappa_1 \\
        - \frac{1}{8 \gamma}[ \partial \Mbar_{1,1} ]
    \Big)
    \frac{\gamma^{2k-1}}{1-6t_4\gamma^2}
    \frac{(2k)!}{(k-1)!k!} .
\end{multline}
The first term comes from the operator insertion at the marked point.
The second arises from the string background, encoded via $T$. The third is supported at the algebraic boundary of $\Mbar_{1,1}$, where the torus degenerates to a pinched one. Using the standard integrals $\int_{\Mbar_{1,1}} \psi_1 = \int_{\Mbar_{1,1}} \kappa_1 = \frac{1}{24}$ and $\int_{\Mbar_{1,1}} [\partial \Mbar_{1,1}] = \frac{1}{2}$, we find
\begin{equation}\label{eq:g1:1pt}
    \corr{\cV_{2k}}{g=1}{quart}
    =
    \biggl(
        \frac{k-1}{12}
        -
        t_4 \gamma^2 \frac{k-2}{2}
    \biggr)
    \frac{\gamma^{2k-2}}{1-12t_4}
    \frac{(2k)!}{(k-1)! k!} .
\end{equation}
One may then check directly that the above equation matches the result derived from topological recursion. As $t_4 \to 0$, one has $\gamma \to 1$, and the formula reduces to the 1986 result of \cite{HZ86} for the genus-one, one-point function of the Gaussian matrix model.

\section{\label{sec:ds:models}Relation to Double-Scaled Models}
How do the well-known dualities between the minimal strings and double-scaled matrix integrals fit into the general correspondence proposed here? 

\smallskip

\textbf{A BMN-like Limit.}
Even \emph{before} taking the double-scaling limit, the observables of the $(2,1)$ minimal string, namely pure topological gravity, already appear as a universal subsector of the string dual to any one-matrix model. Its correlators are given by integrals of $\psi$-classes over moduli space. Physically, this sector is isolated by a BMN-like limit of large traces, that is, by considering operators $\Tr M^k$ with $k \gg 1$. In this regime, the LG matter theory decouples from the topological gravity sector, and the correlators factorize into distinct matter and gravity contributions. For simplicity, we state the result in the one-cut phase for even potentials, where \cref{eq:main:result} gives
\begin{multline} \label{eq:BMN}
    \frac{
        \corr{\prod_{i=1}^n \Tr M^{\kappa a_i}}{g}{}
    }{
    (2\gamma)^{\kappa|a|} \kappa^{3g-3+3n/2}
    }
    \sim
    \left(
        \frac{\corr{\cO_+^n}{g}{LG}}{\Omega_+^n}
        +
        (-1)^{g-1}
        \frac{\corr{\cO_-^n}{g}{LG}}{\Omega_-^n}
    \right) \\
    \times
    \frac{1}{(2\gamma)^{3g-3+n}}
    \sum_{m_1,\ldots,m_n \ge 0}
        \int_{\Mbar_{g,n}}
        \prod_{i=1}^n
        \frac{
            \psi_i^{m_i} a_i^{m_i + \frac{1}{2}}
        }{
            \sqrt{2\pi}
        } .
\end{multline}
The asymptotic is for $\kappa \to \infty$ with $\kappa a_i$ a positive even integer. For the Gaussian model, this limit was derived from the asymptotics of Hurwitz numbers in \cite{OP09}; here we extend the result to arbitrary interacting one-matrix integrals. Physically, this limit probes eigenvalues near the edge of the large-$N$ distribution, where the density vanishes with the universal square-root profile characteristic of Airy universality (cf. \cite{GMM}). \Cref{eq:BMN} is its worldsheet incarnation. In this sense, it plays an analogous role to the pp-wave/BMN limit in AdS/CFT \cite{BMN02}.

\smallskip

\textbf{Double-Scaling Limits on the Worldsheet.}
The matrix-model free energies and correlators diverge as the 't~Hooft couplings approach critical values. For the quartic model, this occurs as $t_4 \to t_{\rm crit} = \frac{1}{12}$, as one can explicitly see from \cref{eq:g0:3pt:exact,eq:g1:1pt}. The double-scaling limit zooms into this critical behavior.

Since our string theory makes sense without double-scaling, we can study this limit directly on the worldsheet. As $t_k \to t_{k,\rm crit}$, the holomorphic top form $\Omega = dy$ develops a zero precisely at the critical points of the superpotential, so the Hessians diverge. Geometrically, the target develops cusps at the critical points. To capture the limiting theory, one must blow up this singularity.

For concreteness, consider the explicit dual of the quartic model. Setting $t_4 = \frac{1-\epsilon^2}{12}$ and $\phi = 1+\sqrt{\frac{\epsilon}{2}} \tilde{\phi} + \bigO(\epsilon)$, one can expand both $W$ and $\Omega$ in powers of $\epsilon$. Keeping the leading terms, after suitable rescalings by fractional powers of $\epsilon$, defines a new LG model with superpotential and holomorphic top form given by
\begin{equation}
    \widetilde{W} = \frac{1}{\sqrt{2}}( \tilde{\phi}^2 - 2),
    \qquad
    \widetilde{\Omega} = (\tilde{\phi}^2 - 1)d\tilde{\phi}.
\end{equation}
The string coupling $g_{\rm s} = 1/N$ is effectively rescaled by a factor of $(t_{\rm crit}-t_4)^{-5/4}$. In the double-scaling limit, we take $g_{\rm s} \to 0$, $t_{4} \to t_{\rm crit}$, while keeping  $\tilde{g}_{\rm s} = g_{\rm s} (t_{\rm crit}-t_{4})^{-5/4}$ fixed. This defines the new genus expansion parameter $\tilde{g}_{s}$. Translating back to matrix-model variables, we recover precisely the spectral curve of the double-scaled model, dual to the $(2,3)$ minimal string \cite{GM90,BK90,DS90,Eyn16}.

Although we have focused on the quartic model, the same geometric mechanism applies more generally. The identification of spectral-curve and B-model data therefore allows one to formulate the double-scaling limit directly at the level of the worldsheet path integral.

\section*{Acknowledgments}
A.G. is supported by an ETH Fellowship (22-2 FEL-003) and a Hermann–Weyl Instructorship from the Forschungsinstitut für Mathematik at ETH Zürich. 
R.G. acknowledges the support of the Department of Atomic Energy, Government of India, under Project No. RTI 4019.
E.A.M. is supported by a SwissMAP Research Fellowship and more broadly by the Swiss National Science Foundation. 
R.G. \& E.A.M. especially thank Matthias Gaberdiel and Wei Li for related collaboration on the A-model string duals to matrix integrals. 
We thank O. Aharony, J. Brödel, K. Costello, F. Coronado, L. Eberhardt, D. Gaiotto, K. Hori, J.H. Lee, W. Lerche, S. Komatsu, M. Mariño, S. Murthy, B. Nairz, K. Osuga, N. Paquette, J. Schmitt, Y. Schuler, S. Shadrin, D. Ranard, V. Rodriguez, and E. Witten for helpful discussions.

\bibliographystyle{apsrev4-2}
\bibliography{letterbib}
    
\appendix
\onecolumngrid
\section{\label{sec:supp}Supplementary Material}
In this appendix, we show how to explicitly compute the worldsheet correlators of vertex operators for the closed-string dual to the quartic matrix model in terms of cohomology classes on $\Mbar_{g,n}$. The resulting expressions are summarized in the \hyperref[sec:checks]{Explicit Cross-Checks} section. The worldsheet data dual to the quartic model (cf. \cref{eq:SC:quart}) are
\begin{equation}
    \cT = \mathbb{CP}^1 \setminus \Set{ 0,\infty, \pm \frac{2}{\gamma^2 \sqrt{12 t_4}} },
    \quad
    W = \gamma \left( \phi + \frac{1}{\phi} \right),
    \quad
    \Omega = \gamma \Bigl( 1 - 3t_4 \gamma^2 \bigl( 1 + \phi^2 \bigr) \Bigr) d\phi,
    \quad
    B(\phi_1,\phi_2) = \frac{d\phi_1 d\phi_2}{(\phi_1 - \phi_2)^2},
\end{equation}
where $\gamma^2 = \frac{1-\sqrt{1-12 t_4}}{6t_4}$.
The point $0$ is a pole of $W$, the point $\infty$ is a pole of both $W$ and $\Omega$, and the points $\pm \frac{2}{\gamma^2 \sqrt{12 t_4}}$ are zeros of $\Omega$; for this reason, they are excised from $\mathbb{CP}^1$. The Hessians $H_\alpha$ and the constants $\Omega_\alpha$, for $\alpha=\pm1$, are
\begin{equation} \label{eq:HOmegt4}
    H_{\alpha}
    \coloneqq
    \left. \frac{\nabla dW}{\Omega^2} \right|_{\alpha}
    =
    \left. \frac{2\gamma \phi^{-3}}{\gamma^2 \bigl( 1 - 3t_4 \gamma^2 (1 + \phi^2) \bigr)^2} \right|_{\alpha}
    =
    \frac{2\alpha}{\gamma (1 - 12 t_4)},
    \quad
    \Omega_{\alpha}
    \coloneqq
    \left. \frac{\Omega}{d\phi} \right|_{\alpha}
    =
    \left. \gamma \Bigl( 1 - 3t_4 \gamma^2 \bigl( 1 + \phi^2 \bigr) \Bigr) \right|_{\alpha}
    =
    \gamma(1 - 6t_4\gamma^2).
\end{equation}
Since the eigenvalue distribution is symmetric for even potentials, the operator dictionary in \cref{eq:op:dict} simplifies further. Expanding the hypergeometric series, we obtain
\begin{equation} \label{eq:evendict}
    \cV_k
    =
    \sum_{\substack{\alpha = \pm 1 \\ m = 0,\ldots, k-1}}
        \cO_\alpha \psi^m
        \frac{\gamma (\alpha \gamma)^{k-m-1}}{\Omega_{\alpha}}
        \frac{k!}{\floor*{\frac{k-m-1}{2}}! \ceil*{\frac{k-m-1}{2}}!} .
\end{equation}
To compute the integrand $\corrint{\prod_i \cV_{k_i}}{g}{LG+grav}$, one first expands each vertex operator $\cV_{k_i}$ linearly in the basis $\cO_\alpha \psi^m$ and pulls the explicit $\psi$-classes outside the gravitationally dressed matter correlator. One is then left with the problem of computing $\corrint{\prod_i \cO_{\alpha_i}}{g}{LG+grav}$, which, by definition, is the right-hand side of \cref{eq:corr:LGgrav}: the cohomology class obtained by applying the data $T$ and $R$ to the pure matter correlators. We now explain how this works in practice, and how to interpret it from the worldsheet point of view.

\smallskip

\textbf{Genus-0, 3-point.}
For $(g,n)=(0,3)$, only the matter sector contributes, since the coupling to topological gravity is trivial: $\corrint{\prod_{i=1}^3 \cO_{\alpha_i}}{g=0}{LG+grav} = \corr{\prod_{i=1}^3 \cO_{\alpha_i}}{g=0}{LG}$. Moreover, since $\psi$ represents a two-form and $\Mbar_{0,3}$ is a point, only the $m=0$ term in \cref{eq:evendict} contributes. Hence:
\begin{equation}
    \Corrint{
        \prod_{i=1}^3 \cV_{2k_i}
    }{g=0}{LG+grav}
    =
    \sum_{\alpha_i = \pm 1} \corr{
        \cO_{\alpha_1} \cO_{\alpha_2} \cO_{\alpha_3}
    }{g=0}{LG}
    \prod_{i=1}^{3}
        \frac{\alpha_i \gamma^{2k_i}}{\Omega_{\alpha_i}}
        \frac{(2k_i)!}{(k_i-1)!k_i!}
    =
    \left(
        \frac{(+1)^3}{H_+ \Omega_+^3}
        +
        \frac{(-1)^3}{H_- \Omega_-^3}
    \right)
    \prod_{i=1}^3 \frac{\gamma^{2k_i}(2k_i)!}{(k_i-1)!k_i!}
    .
\end{equation}
In the last step we used the diagonality of the pure matter correlators, cf. \cref{eq:LG:diag:corr}. Since there is no moduli-space integral to perform, this is already the full answer. Substituting \cref{eq:HOmegt4} reproduces \cref{eq:g0:3pt:exact}.

\smallskip

\textbf{Genus-1, 1-point \& a Quick Physical Primer on (Semisimple) CohFTs.}
The case $(g,n) = (1,1)$ provides the first non-trivial test of the coupling to topological gravity. In particular, the computation of $\corrint{\cO_{\alpha}}{g=1}{LG+grav}$ illustrates in practice how the data of $T$ and $R$ reconstruct the gravity-coupled theory from the pure matter sector.

\begin{figure}[b]
    \centering
    \begin{subfigure}[b]{0.28\columnwidth}
        \begin{tikzpicture}[x=1pt,y=1pt,scale=.6]
            \fill[opacity=.05](186.667, 568) .. controls (181.333, 557.333) and (170.667, 546.667) .. (154.667, 541.333) .. controls (138.667, 536) and (117.333, 536) .. (101.333, 541.333) .. controls (137.7777, 545.7777) and (166.2223, 554.6667) .. (186.667, 568) -- cycle;
            \draw(96, 600) .. controls (96, 568) and (160, 568) .. (160, 600);
            \draw(200.077, 563.242) -- (200.077, 563.242);
            \node at (88, 564) {$\bullet$};
            \node at (128, 616) {$\bullet$};
            \draw(103.369, 583.998) .. controls (108, 596) and (116, 598) .. (121.2415, 602.3417) .. controls (126.483, 606.6833) and (128.966, 613.3667) .. (127.6327, 618.7) .. controls (126.2993, 624.0333) and (121.1497, 628.0167) .. (114.5748, 630.0083) .. controls (108, 632) and (100, 632) .. (91.3333, 628) .. controls (82.6667, 624) and (73.3333, 616) .. (68.6667, 605.3333) .. controls (64, 594.6667) and (64, 581.3333) .. (69.3333, 569.3333) .. controls (74.6667, 557.3333) and (85.3333, 546.6667) .. (101.3333, 541.3333) .. controls (117.3333, 536) and (138.6667, 536) .. (154.6667, 541.3333) .. controls (170.6667, 546.6667) and (181.3333, 557.3333) .. (186.6667, 569.3333) .. controls (192, 581.3333) and (192, 594.6667) .. (187.3333, 605.3333) .. controls (182.6667, 616) and (173.3333, 624) .. (164.6667, 628) .. controls (156, 632) and (148, 632) .. (141.428, 629.9987) .. controls (134.856, 627.9973) and (129.712, 623.9947) .. (128.3787, 618.6613) .. controls (127.0453, 613.328) and (129.5227, 606.664) .. (134.7613, 602.332) .. controls (140, 598) and (148, 596) .. (152.629, 583.996);
            \fill[opacity=.05](186.667, 688) .. controls (181.333, 677.333) and (170.667, 666.667) .. (154.667, 661.333) .. controls (138.667, 656) and (117.333, 656) .. (101.333, 661.333) .. controls (137.7777, 665.7777) and (166.2223, 674.6667) .. (186.667, 688) -- cycle;
            \draw(69.3333, 688) .. controls (74.6667, 677.3333) and (85.3333, 666.6667) .. (101.3333, 661.3333) .. controls (117.3333, 656) and (138.6667, 656) .. (154.6667, 661.3333) .. controls (170.6667, 666.6667) and (181.3333, 677.3333) .. (186.6667, 688) .. controls (192, 698.6667) and (192, 709.3333) .. (186.6667, 720) .. controls (181.3333, 730.6667) and (170.6667, 741.3333) .. (154.6667, 746.6667) .. controls (138.6667, 752) and (117.3333, 752) .. (101.3333, 746.6667) .. controls (85.3333, 741.3333) and (74.6667, 730.6667) .. (69.3333, 720) .. controls (64, 709.3333) and (64, 698.6667) .. cycle;
            \draw(96, 720) .. controls (96, 688) and (160, 688) .. (160, 720);
            \draw(152.629, 703.996) .. controls (144, 724) and (112, 724) .. (103.369, 703.998);
            \draw(200.077, 683.242) -- (200.077, 683.242);
            \node at (88, 684) {$\bullet$};
        \end{tikzpicture}
        \caption{A once-punctured torus (top) and its possible degeneration to a pinched torus (bottom).}
        \label{fig:11:stbl:graphs}
    \end{subfigure}
    \hfill
    \begin{subfigure}[b]{0.34\columnwidth}
        \begin{tikzpicture}[x=1pt,y=1pt,scale=.75]
            \fill[opacity=.05](346.667, 688) .. controls (341.333, 677.333) and (330.667, 666.667) .. (314.667, 661.333) .. controls (298.667, 656) and (277.333, 656) .. (261.333, 661.333) .. controls (297.7777, 665.7777) and (326.2223, 674.6667) .. (346.667, 688) -- cycle;
            \filldraw[fill opacity=.2](260.7817, 691.7272) .. controls (257.5518, 696.8047) and (249.1916, 701.6479) .. (245.202, 699.0783) .. controls (241.2125, 696.5088) and (241.5936, 686.5265) .. (244.8235, 681.449) .. controls (248.0534, 676.3716) and (254.132, 676.1988) .. (258.1215, 678.7684) .. controls (262.1111, 681.3379) and (264.0116, 686.6498) .. cycle;
            \filldraw[fill opacity=.2](280.066, 747.682) .. controls (275.978, 749.545) and (273.989, 746.7725) .. (272.9945, 742.7196) .. controls (272, 738.6667) and (272, 733.3333) .. (273.3918, 728.6069) .. controls (274.7835, 723.8805) and (277.567, 719.761) .. (280.024, 721.404) .. controls (284, 724) and (286, 730) .. (285.9975, 735.3276) .. controls (285.9951, 740.6552) and (283.9901, 745.3103) .. (280.2195, 747.5939);
            \filldraw[fill opacity=.2](296.028, 747.712) .. controls (299.817, 749.457) and (301.9085, 746.7285) .. (302.9543, 742.6976) .. controls (304, 738.6667) and (304, 733.3333) .. (302.5243, 728.6597) .. controls (301.0485, 723.986) and (298.097, 719.972) .. (295.994, 721.384) .. controls (292, 724) and (290, 730) .. (290.009, 735.3267) .. controls (290.0181, 740.6534) and (292.0361, 745.3067) .. (295.874, 747.6241);
            \draw[opacity=.3](280.024, 761.404) .. controls (281.7405, 762.6966) and (283.1988, 769.2867) .. (283.3517, 775.1558) .. controls (283.5045, 781.0249) and (282.3519, 786.173) .. (280.219, 787.594);
            \draw[opacity=.3](295.994, 761.384) .. controls (294.2635, 762.6937) and (293.1318, 769.3469) .. (293.0691, 775.3278) .. controls (293.0065, 781.3088) and (294.013, 786.6176) .. (295.874, 787.624);
            \draw(256, 720) .. controls (256, 688) and (320, 688) .. (320, 720);
            \draw(360.077, 683.242) -- (360.077, 683.242);
            \node at (288, 776) {$\bullet$};
            \node at (222.5362, 663.5622) {$\bullet$};
            \draw(235.3178, 671.2897) .. controls (232.0878, 676.3672) and (223.7275, 681.2102) .. (219.738, 678.6406) .. controls (215.7485, 676.0709) and (216.1298, 666.0886) .. (219.3598, 661.0111) .. controls (222.5898, 655.9336) and (228.6685, 655.7609) .. (232.658, 658.3306) .. controls (236.6475, 660.9002) and (238.5478, 666.2122) .. cycle;
            \draw(263.369, 703.998) .. controls (268, 716) and (276, 718) .. (280.0243, 721.4035);
            \draw(280.2195, 747.5939) .. controls (278.5234, 748.5782) and (276.6179, 749.3894) .. (274.5748, 750.0083) .. controls (268, 752) and (260, 752) .. (251.3333, 748) .. controls (242.6667, 744) and (233.3333, 736) .. (228.6667, 725.3333) .. controls (224, 714.6667) and (224, 701.3333) .. (229.3333, 689.3333) .. controls (234.6667, 677.3333) and (245.3333, 666.6667) .. (261.3333, 661.3333) .. controls (277.3333, 656) and (298.6667, 656) .. (314.6667, 661.3333) .. controls (330.6667, 666.6667) and (341.3333, 677.3333) .. (346.6667, 689.3333) .. controls (352, 701.3333) and (352, 714.6667) .. (347.3333, 725.3333) .. controls (342.6667, 736) and (333.3333, 744) .. (324.6667, 748) .. controls (316, 752) and (308, 752) .. (301.428, 749.9987) .. controls (299.4209, 749.3874) and (297.5469, 748.5895) .. (295.874, 747.6241);
            \draw(295.8978, 721.4539) .. controls (301.1015, 717.6717) and (308.3234, 715.1613) .. (312.629, 703.996);
            \draw(296.0278, 787.7121) .. controls (299.8172, 789.4574) and (301.9086, 786.7287) .. (302.9543, 782.6977) .. controls (304, 778.6667) and (304, 773.3333) .. (302.5241, 768.6596) .. controls (301.0483, 763.9858) and (298.0966, 759.9716) .. (295.8978, 761.4539) .. controls (289.5227, 766.664) and (287.0453, 773.328) .. (288.3787, 778.6613) .. controls (289.3048, 782.3658) and (292.0694, 785.4283) .. (296.0278, 787.7121);
            \draw(280.066, 787.6822) .. controls (275.9776, 789.5454) and (273.9888, 786.7727) .. (272.9944, 782.7197) .. controls (272, 778.6667) and (272, 773.3333) .. (273.3919, 768.607) .. controls (274.7837, 763.8807) and (277.5674, 759.7613) .. (280.0243, 761.4035) .. controls (286.483, 766.6833) and (288.966, 773.3667) .. (287.6327, 778.7) .. controls (286.7137, 782.376) and (283.9816, 785.4107) .. (280.066, 787.6822);
            \node at (238, 700) {$\beta$};
            \node at (265, 740) {$\mu$};
            \node at (311, 740) {$\nu$};
            \node at (205, 690) {$R_\alpha^\beta(\psi_1)$};
            \node at (345, 776) {$E^{\mu,\nu}(\psi',\psi'')$};
            \node at (218, 653) {$\mathcal{O}_{\alpha}$};
        \end{tikzpicture}
        \caption{The $R$-action: $E^{\mu,\nu}$ encodes the sewing of boundary states $\ket{\mu}$ and $\ket{\nu}$ across a node, while $R^{\beta}_{\alpha}$ corresponds to a disk partition function with an $\cO_{\alpha}$ insertion and boundary state $\ket{\beta}$.}
        \label{fig:R:action}
    \end{subfigure}
    \hfill
    \begin{subfigure}[b]{0.34\columnwidth}
        \begin{tikzpicture}[x=1pt,y=1pt,scale=.75]
            \fill[opacity=.05](186.667, 688) .. controls (181.333, 677.333) and (170.667, 666.667) .. (154.667, 661.333) .. controls (138.667, 656) and (117.333, 656) .. (101.333, 661.333) .. controls (137.7777, 665.7777) and (166.2223, 674.6667) .. (186.667, 688) -- cycle;
            \filldraw[fill opacity=.2](182.6667, 717.3333) .. controls (181.3333, 722.6667) and (174.6667, 729.3333) .. (170.6667, 728.6667) .. controls (166.6667, 728) and (165.3333, 720) .. (166.6667, 714.6667) .. controls (168, 709.3333) and (172, 706.6667) .. (176, 707.3333) .. controls (180, 708) and (184, 712) .. cycle;
            \filldraw[fill opacity=.2](100.7817, 691.7275) .. controls (97.5517, 696.805) and (89.1913, 701.648) .. (85.2018, 699.0783) .. controls (81.2123, 696.5087) and (81.5937, 686.5263) .. (84.8237, 681.4488) .. controls (88.0537, 676.3713) and (94.1323, 676.1987) .. (98.1218, 678.7683) .. controls (102.1113, 681.338) and (104.0117, 686.65) .. cycle;
            \draw(69.3333, 688) .. controls (74.6667, 677.3333) and (85.3333, 666.6667) .. (101.3333, 661.3333) .. controls (117.3333, 656) and (138.6667, 656) .. (154.6667, 661.3333) .. controls (170.6667, 666.6667) and (181.3333, 677.3333) .. (186.6667, 688) .. controls (192, 698.6667) and (192, 709.3333) .. (186.6667, 720) .. controls (181.3333, 730.6667) and (170.6667, 741.3333) .. (154.6667, 746.6667) .. controls (138.6667, 752) and (117.3333, 752) .. (101.3333, 746.6667) .. controls (85.3333, 741.3333) and (74.6667, 730.6667) .. (69.3333, 720) .. controls (64, 709.3333) and (64, 698.6667) .. cycle;
            \draw(96, 720) .. controls (96, 688) and (160, 688) .. (160, 720);
            \draw(152.629, 703.996) .. controls (144, 724) and (112, 724) .. (103.369, 703.998);
            \draw(200.077, 683.242) -- (200.077, 683.242);
            
            \node at (204, 728) {$\bullet$};
            \draw(210.6667, 729.3333) .. controls (209.3333, 734.6667) and (202.6667, 741.3333) .. (198.6667, 740.6667) .. controls (194.6667, 740) and (193.3333, 732) .. (194.6667, 726.6667) .. controls (196, 721.3333) and (200, 718.6667) .. (204, 719.3333) .. controls (208, 720) and (212, 724) .. cycle;
            \node at (165, 732) {$\sigma$};
            \node at (185, 752) {$T^\sigma(\psi_2)$};
            \node at (62.536, 663.5622) {$\bullet$};
            \draw(75.3179, 671.2897) .. controls (72.0879, 676.3672) and (63.7275, 681.2102) .. (59.738, 678.6405) .. controls (55.7485, 676.0709) and (56.1299, 666.0885) .. (59.3599, 661.011) .. controls (62.5899, 655.9335) and (68.6685, 655.7609) .. (72.658, 658.3305) .. controls (76.6475, 660.9002) and (78.5479, 666.2122) .. cycle;
            \node at (78, 700) {$\beta$};
            \node at (45, 690) {$R_\alpha^\beta(\psi_1)$};
            \node at (58, 653) {$\mathcal{O}_{\alpha}$};
            \node at (240, 656) {$\vphantom{\beta}$};
        \end{tikzpicture}
        \caption{The $T$-action describes how the choice of background is realized by integrating out descendant insertions over internal marked points on the worldsheet labeled by internal boundary states.}
        \label{fig:T:action}
    \end{subfigure}
    \caption{A pictorial representation of the worldsheet Feynman calculus computing semisimple CohFT correlators.}
    \label{fig:WS:Feynman:calculus}
\end{figure}
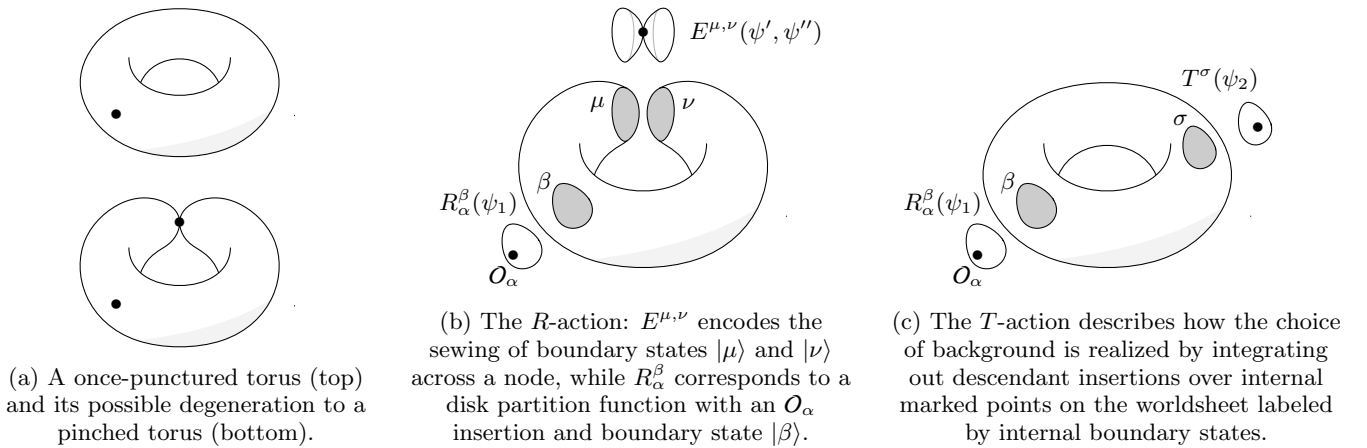

A useful physical picture for the semisimple CohFTs relevant here is as a worldsheet analog of a Feynman-diagram expansion. One starts from a genus-$g$ worldsheet with $n$ marked points corresponding to external operator insertions, and considers all allowed degeneration limits of that surface. In the mathematics literature, these are encoded by stable graphs \cite{GL26}. Physically, they describe ways in which cycles on the worldsheet pinch and operator insertions collide, so that the surface breaks into simpler components joined at nodes. Stability means that every component of genus $g_c$ with $n_c$ special points---that is, marked points or branches of nodes---must satisfy $2g_c - 2 + n_c > 0$.

For $(g,n) = (1,1)$, there are only two relevant configurations, see \cref{fig:11:stbl:graphs}. The first is the smooth torus with one marked point. The second is its unique stable degeneration: one cycle pinches, producing a thrice-punctured sphere in which two punctures are glued together to form a node, while the third remains the marked point. This reflects the well-known identification $\Mbar_{1,1} \setminus \M_{1,1} \cong \Mbar_{0,3}/\mathbb{Z}_2$ (the $\mathbb{Z}_2$ represents the residual automorphism that exchanges the two branches of the node).

We now describe the $R$-action. To this end, we excise a small disk around each marked point and around each branch of every node, as in \cref{fig:R:action}. In this way, both operator insertions and nodes are described by gluing disks to the remaining surface, while summing over their allowed boundary conditions. Concretely, if a marked point carries the insertion $\cO_{\alpha_i}$, then excising a disk around it produces a disk one-point function with boundary state labeled by some $\beta$. This contributes a factor of $R_{\alpha_i}^\beta(\psi_i)$. As we will show in \cite{GGM}, this is precisely the disk one-point function with bulk insertion $\cO_{\alpha_i}$ and boundary state $\ket{\beta}$. Similarly, when a node is formed, one excises disks around its two branches. This gives the edge term $E^{\mu,\nu}(\psi',\psi'')$, where $\mu$ and $\nu$ label the boundary states associated with the two branches of the node, and $\psi',\psi''$ are the corresponding gravitational descendants. Physically, $E^{\mu,\nu}$ may likewise be viewed as a degenerate cylinder amplitude, itself expressed in terms of simpler disk partition functions:
\begin{equation} \label{eq:edgeR}
    E^{\mu,\nu}(u',u'')
    \coloneqq
    \frac{\eta^{\mu,\nu} - R^\mu_\rho(u') \eta^{\rho,\sigma}R^\nu_\sigma(u'')}{u'+u''},
\end{equation}
where $\eta_{\mu,\nu} \coloneqq \corr{\cO_\mu \cO_\nu}{g=0}{LG}$ is the pure-matter cylinder amplitude and $\eta^{\mu,\nu}$ its inverse.

The $T$-action, instead, accounts for how the choice of background is reflected in the coupling of the worldsheet theory to topological gravity. It introduces additional \emph{internal} marked points at which $\psi$-classes are inserted. One must integrate over the positions of these internal points on the worldsheet. In this sense, the $T$-action describes a gas of integrated background insertions spread over the surface. Mathematically, this is implemented by introducing extra marked points and then pushing forward along the forgetful map that removes them; this pushforward is precisely the integration over their positions on the worldsheet. The internal marked points may again be represented by excising small disks around them and summing over the corresponding boundary labels; see \cref{fig:T:action}. Each insertion of $T$ is thus also labeled by the boundary state of the disk, $\ket{\sigma}$, and contributes a factor of $T^{\sigma}(\psi)$. Only finitely many $T$-insertions can contribute for fixed $g$ and $n$: once too many descendants are inserted, the total degree exceeds the dimension of the moduli space, and the corresponding contribution vanishes.

Why excise disks? The key observation of \cite{VV91} is that the effect of topological gravity can be localized at the marked points and nodes of the worldsheet. Once small disks around these special points are removed, the remaining surface behaves as if worldsheet gravity were turned off, so that each component contributes only through pure topological field theory. All non-trivial gravitational corrections are then encoded in the disk partition functions \cref{eq:T,eq:R}. This is our proposed worldsheet interpretation of \cref{eq:corr:LGgrav}.

We now have all the ingredients needed to compute $\corrint{\cO_\alpha}{g=1}{LG+grav}$. The general expression is
\begin{equation}\label{eq:g11-gen}
\begin{split}
    \corrint{\cO_\alpha}{g=1}{LG+grav}
    = 
    \sum_{\beta}
        R_\alpha^\beta(\psi_1)
        \biggl(
            \corr{\cO_\beta}{g=1}{LG}
            +
            \sum_{\sigma}
                \corr{\cO_\beta \cO_{\sigma}}{g=1}{LG}
                \pi_{*} T^\sigma(\psi_{2})
            + \underbrace{\cdots}_{=0}
        \biggr) & \\
    +
    \frac{1}{2}
    \iota_*
    \sum_{\beta,\mu,\nu}
    R_\alpha^\beta(\psi_1)\,
    E^{\mu,\nu}(\psi',\psi'')
    \biggl(
        \corr{\cO_\beta\cO_\mu\cO_\nu}{g=0}{LG}
        +
        \underbrace{\cdots}_{=0}
    \biggr) & .
\end{split}
\end{equation}
This formula applies to any LG model coupled to gravity of the type considered in this letter and, in particular, to the string dual of any interacting one-matrix model. The two terms in \cref{eq:g11-gen} are in one-to-one correspondence with the worldsheet configurations shown in \cref{fig:11:stbl:graphs}. Here, $\pi \colon \Mbar_{1,2} \to \Mbar_{1,1}$ is the forgetful map that removes the additional marked point associated with the insertion of $T$, and $\pi_*$ realizes the integration over the possible position of that internal marked point on the worldsheet. Moreover, $ \iota \colon \Mbar_{0,3} \to \Mbar_{1,1}$ is the gluing map that identifies the two branches of the node, thereby realizing the thrice-punctured sphere as the algebraic boundary of $\Mbar_{1,1}$. The symmetry factor of $1/2$ accounts for the exchange of the two branches of the node. The ellipses stand for terms with additional internal marked points, each weighted by a further factor of $T$. For $(g,n)=(1,1)$, however, these higher terms do not contribute: every extra insertion raises the degree of the integrand, and one quickly exceeds the dimension of the relevant moduli space. Thus, in this example, the expansion truncates after the terms displayed above.

On dimensional grounds, \cref{eq:g11-gen} simplifies even further. In the following, set $[f]_k$ for the coefficient of $u^k$ in the Taylor expansion of $f(u)$. In the first line, only the constant $[R_\alpha^\beta]_0$ and the linear term $[R_\alpha^\beta]_1 \psi_1$ in the expansion of $R$ can contribute. The $T$-piece similarly truncates to the first non-trivial term, $[T^\sigma]_2 \psi_2^2$. Integrating along the position of the additional insertion, this can be written on $\Mbar_{1,1}$ in terms of $\kappa$-classes: $\pi_* \psi_2^2 = \kappa_1$. In the second line, only the degree-zero part of $R$ and the edge factor survive, $[R_\alpha^\beta]_0$ and $[E^{\mu,\nu}]_{0,0}$, since $\Mbar_{0,3}$ is a point. The term $\frac{1}{2} \iota_* 1$ is often denoted as $[\partial \Mbar_{1,1}]$, since it coincides with the Poincaré dual of the algebraic boundary. Expanding to the relevant order therefore gives
\begin{equation}\label{eq:g11-class-expanded}
\begin{split}
    \corrint{\cO_\alpha}{g=1}{LG+grav}
    =
    \sum_{\beta}
    \Bigl(
        \bigl[R_\alpha^\beta\bigr]_0
        +
        \bigl[R_\alpha^\beta\bigr]_1
        \psi_1
    \Bigr)
    \corr{\cO_\beta}{g=1}{LG}
    +
    \sum_{\beta,\sigma}
        \bigl[R_\alpha^\beta\bigr]_0
        \bigl[T^\sigma\bigr]_2
        \corr{\cO_\beta \cO_{\sigma}}{g=1}{LG} 
        \kappa_1
    & \\
    + 
    \sum_{\beta,\mu,\nu}
        \bigl[R_\alpha^\beta\bigr]_{0}
        \bigl[E^{\mu,\nu}\bigr]_{0,0}
        \corr{\cO_\beta\cO_\mu\cO_\nu}{g=0}{LG}
        \left[ \partial \Mbar_{1,1} \right]
    &.
\end{split}
\end{equation}
Again, this expression is completely general: it applies to the string dual of any one-matrix model. As promised, it is a sum of classes on $\Mbar_{1,1}$, namely $1$, $\psi_1$, $\kappa_1$, and $[\partial \Mbar_{1,1}]$, with coefficients determined by $R$, $T$, and the pure matter correlators. It is the specialization of the right-hand side of \cref{eq:corr:LGgrav} to the case $(g,n) = (1,1)$. All that remains is to substitute the Taylor expansions of $R$, $T$, and the edge term $E$.

In the one-cut phase, for any choice of even matrix-model potential, the matrix-valued function $R$ in \cref{eq:R} is
\begin{equation}
\begin{split}
    R^\alpha_\alpha(u)
    =
    {}_2F_0\bigl(
        \tfrac{1}{2},-\tfrac{1}{2};; -\tfrac{u}{4\alpha\gamma}
    \bigr)
    =
    1
    +\frac{\alpha}{16\gamma} u
    +\cdots
    ,
    \qquad
    R^\alpha_{-\alpha}(u)
    =
    \frac{u}{8 \alpha \gamma} \,
    {}_2F_0\bigl(
      \tfrac{3}{2},\tfrac{1}{2};; -\tfrac{u}{4\alpha\gamma}
    \bigr)
    =
    \frac{\alpha}{8\gamma} u
    +
    \cdots,
\end{split}
\end{equation}
where ${}_p F_q$ denotes the generalized hypergeometric series. The edge term is then obtained from \cref{eq:edgeR}. For the worldsheet dual to the quartic model, its first Taylor coefficients are
\begin{equation} \label{eq:edge}
    E^{\alpha,\alpha}(u',u'')
    =
    - \tfrac{1}{8\gamma^2(1-12 t_4)}
    + \tfrac{3\alpha(u'+u'')}{256\gamma^3(1-12t_4)}
    + \cdots ,
    \qquad
    E^{\alpha,-\alpha}(u',u'')
    =
    \tfrac{1}{4\gamma^2 (1-12 t_4)}
    - \tfrac{3\alpha(u'-u'')}{64\gamma^3(1-12 t_4)}
    + \cdots .
\end{equation}
The exponential integral defining $T$ in \cref{eq:T} can also be evaluated explicitly. For the quartic model, one finds
\begin{equation}
    T^\alpha(u)
    =
    u
    \Bigl(
    1
    -
    {}_2 F_0\bigl(
        \tfrac{3}{2},-\tfrac{1}{2};;-\tfrac{u}{4\alpha\gamma}
    \bigr)
    +
    \frac{6t_4 \, \alpha\gamma u}{1-6t_4 \gamma^2} \,
    {}_2 F_0\bigl(
        \tfrac{5}{2},-\tfrac{3}{2};;-\tfrac{u}{4\alpha\gamma}
    \bigr)
    \Bigr)
    =
    - \frac{3\alpha(1 - 38 t_4\gamma^2)}{16\gamma(1 - 6 t_4\gamma^2)} u^2
    +
    \cdots .
\end{equation}
Substituting the corresponding coefficients into \cref{eq:g11-class-expanded}, we obtain
\begin{equation} \label{eq:Oquartic}
    \corrint{\cO_\alpha}{g=1}{LG+grav} \Big|_{\textrm{quartic}}
    = 
    1
    - \frac{\alpha}{16 \gamma} \psi_1
    - \frac{3\alpha(1-38t_4\gamma^2)}{16\gamma(1-6 t_4\gamma^2)} \kappa_1
    - \frac{\alpha}{16 \gamma} [\partial \Mbar_{1,1}].
\end{equation}
To compute the first $1/N$ correction to the one-point function of the quartic matrix model, it remains to evaluate $\corrint{\cV_{2k}}{g=1}{LG+grav}$. Since $\Mbar_{1,1}$ has complex dimension $1$, we need only the terms up to $m=1$ in \cref{eq:evendict}.
Using \cref{eq:Oquartic} and retaining only terms of cohomological degree at most $2$, we find the explicit integrand on moduli space dual to $\corr{\Tr M^{2k}}{g=1}{quart}$, as quoted in the main text:
\begin{equation}\label{eq:g11-vertex}
    \corrint{\cV_{2k}}{g=1}{LG+grav} \Big|_{\textrm{quartic}}
    =
    \left(
        \frac{16k - 1}{8\gamma} \psi_1
        - \frac{3(1-38t_4\gamma^2)}{8\gamma(1-6 t_4\gamma^2)} \kappa_1
        - \frac{1}{8 \gamma} [\partial \Mbar_{1,1}]
    \right)
    \frac{\gamma^{2k-1}}{1-6t_4 \gamma^2}
    \frac{(2k)!}{(k-1)!k!}.
\end{equation}
Integrating over $\Mbar_{1,1}$ gives precisely \cref{eq:g1:1pt}.

\end{document}